\documentstyle[epsf,preprint,aps]{revtex}
\tightenlines
\def\Journal#1#2#3#4{{#1} {\bf #2}, #3 (#4)}
\def\ADP{{\em Adv. in Phys.}}

\def\APG{{\em Ann. Physik}}
\def\CMP{{\em Commun. Math. Phys.}}

\def\JETP{{\em Sov. Phys. J. JETP}}

\def\JPF{{\em J. Phys.} F}

\def\PRB{{\em Phys. Rev.} B}

\def\ZPB{{\em Z. Phys.} B}

\newcommand{\be}{\begin{equation}}
\newcommand{\ee}{\end{equation}}
\newcommand{\bea}{\begin{eqnarray}}
\newcommand{\eea}{\end{eqnarray}}

\newcommand{\hf} {{1\over2}}
\newcommand{\nonu}{\nonumber\\}
\def\la{\langle}
\def\ra{\rangle}
\def\ord{{\cal O}}
\def\eq#1{(\ref{#1})}

\def\hphi{\hat\phi}
\def\hpsi{\hat\psi}
\def\hrho{\hat\rho}
\def\hx{{\hat x}}
\def\hti{\hat t}

\begin{document}
\title{Modelling static impurities}

\author{Sebastiao Correia$^a$\thanks{correia@lpt1.u-strasbg.fr},
Janos Polonyi$^{ab}$\thanks{polonyi@fresnel.u-strasbg.fr},
Jean Richert$^a$\thanks{richert@lpt1.u-strasbg.fr}}
\address{$^a$Laboratoire de Physique Th\'eorique, Universit\'e
Louis Pasteur\\
3 rue de l'Universit\'e 67084 Strasbourg, Cedex, France}
\address{$^b$Department of Atomic Physics, L. E\"otv\"os University\\
P\'azm\'any P. S\'et\'any 1/A 1117 Budapest, Hungary}

\date{\today}
\maketitle
\begin{abstract}
A simple model is presented for the calculation of the quenched
average over impurities which are rendered static by setting
their mass equal to infinity. The path integral formalism
of the second quantized theory contains annealed averages only.
The similarity with the Gaussian quenched potential model is discussed.
\end{abstract}

\section{Introduction}
Disordered systems can be characterized by two time scales,
the time $t$ spent between the preparation of the
impurity distribution within the sample and the measurement,
and the impurity diffusion time, $t_i$. There is actually a third
characteristic time, the time span of the measurement, $t_m$ but
it is far from the other time scales, $t_m<<t,t_i$. We are interested in
the dynamics in the limit when $t,t_i\to\infty$ but
$R=t/t_i$ stays finite. The initial distribution of the
impurities within a given sample determined by the preparation remains
unchanged when $R<<1$. For $R>>1$ the impurities reach equilibrium
with the faster degrees of freedom, electrons. The self averaging quantities
measured on large samples can be obtained by means of
averaging over the impurity distributions. The latter, the
impurity distribution is determined by the preparation
method when $R<<1$ or by the equilibrium properties for $R>>1$.

The usual modelisation methods were developed for $R<<1$. The
time available for the impurity motion in this case is
insufficient to establish the feedback of the fast degrees of
freedom on the impurity dynamics. Such a suppression of
a part of the dynamics can be realized by the replica method
\cite{repl}, \cite{wegn} the introduction of spurious supersymmetric
particles \cite{susy} or the Keldysh formalism \cite{keld}.
A simple model to describe electrons in the presence of quenched
disorder is based on the partition function \cite{wegn}
\be\label{edw}
Z_n=\int D[v]D[\psi^\dagger]D[\psi]e^{-\hf\int dxv^2(x)
+i\sum_j\int dxdt[\psi^\dagger_j(x,t)
(i\partial_t+{\hbar\over2m}\Delta-gv(x))\psi_j(x,t)]},
\ee
where $v(x)$ is a static potential representing the disorder,
$\psi_j$ denotes the electron field, and $j=1,\cdots,n$ is the replica
index. The limit $n\to0$ is made formally at the end of the computation
of the observables.

Although these methods have already led to a number of important results
they are not completely satisfactory.
The strategy based on supersymmetry can not cope with the
Coulomb interactions. The Keldysh formalism, given in terms
of the propagators, is inherently perturbative. The replica
method is not always reliable due to its formal nature.

Apart from glassy materials or fast quenching, the usual preparation 
of samples gives $R>>1$ and there is time for the electron dynamics to
influence the impurity distribution seen during the measurement.
One can easily incorporate the feedback of the electron dynamics
into the impurity distribution by avoiding the replica method in
the model \eq{edw}, i.e. by setting $n=1$. But the simple Gaussian
distributed static field which is sufficient for the modelisation
of the forces due to decoupled impurities might be too simple for the
reproduction of the full impurity dynamics with its non-trivial 
time dependence.

We describe in this paper a simple model for $R>>1$, when the slowly
moving impurities are in equilibrium with the electrons.
Instead of the static potential $v(x)$ and the
replica method used in \eq{edw}, we return to the more detailed impurity
dynamics. The impurities will be made static as far as the observables
are concerned, by setting their mass equal to infinity. Their
equilibrium with the electrons will be established by
considering the partition function at finite density or temperature.
Our model involves particle degrees of freedom, the impurities
and the electrons. The elimination of the former produces an effective
model similar to \eq{edw} with $n=1$.
We find that the salient feature of quenched disorder,
the possibility of generating localisation, is realised in the same
manner for $R>>1$ as for $R<<1$.
In particular, the perturbative solution of our model for the
electronic observables obtained by resumming the two-loop self
energy can be mapped onto a subset of graphs coming from
\eq{edw}, namely those graphs which have no more than two impurity lines
ending at the same space location (our impurities obey
fermion statistics). This result implies that the conductivity,
when computed through the resummation of the maximally crossed diagramms
in the particle-hole channel, agrees in the two models. Since
our model is given in terms of an annealed partition function this result
opens the way for the application of different non-perturbative
numerical methods to deal with quenched disorder when $R>>1$.

The organization of the paper is the following.
Section II introduces our model which contains impurities
with infinite mass, i.e. zero mobility. The infinite 
mass non-relativistic particles have a singular Fermi sphere,
reflecting the high level of degeneracy of the ground state.
This makes the use of the grand canonical ensemble questionable
in this case. The canonical ensemble is not spoiled by the
degeneracy of the free Hamiltonian. In section III
we present a method to compute averages in this ensemble.
The propagators are computed in
section IV and section V describes the setting up of the
perturbation expansion in the framework of the path integral
formalism. The self energy and the propagator for a particle-hole
pair are computed in section VII.
Finally conclusions are drawn in section VIII.
Technical points concerning the regularization and the
perturbation expansion are developed in the Appendices.

\section{The model}\label{model}
Consider the system of electrons and impurities described by
the fields $\psi$ and $\phi$, respectively. The hamiltonian is of the form
\bea\label{mham}
H&=&\int dx\biggl[-{\hbar^2\over2m}
\psi^\dagger_a(x)\Delta\psi_a(x)
+{\lambda\over4}\psi^\dagger_a(x)\psi_a(x)
\psi^\dagger_b(x)\psi_b(x)\nonu
&&-{\hbar^2\over2M}\phi^\dagger(x)\Delta\phi(x)
+g\psi^\dagger_a(x)\psi_a(x)\phi^\dagger(x)\phi(x)\biggr],
\eea
where $a$ and $b$ stand for the spin indices. Electrons and
impurities are present with finite density, what is usually
achieved by using the grand canonical ensemble, the modification of the
Hamiltonian,
\be\label{gcham}
H\to H-\int dx\left[\mu_e\psi^\dagger_a\psi_a(x)
+\mu_i\phi^\dagger(x)\phi(x)\right].
\ee
Here $\mu_e$ and $\mu_i$ are the chemical potential for the electrons
and the impurities, respectively. The spinless impurity field is
supposed to be anticommuting, a reasonable approximation even for
integer spin atoms in solids. In Eq. \eq{gcham} $H$ is considered as
an effective hamiltonian for the conducting band with a
spatial resolution which is so low that the inhomogeneity of the
crystalline structure is not seen. 

We render the impurities static by setting $M=\infty$. Then the lagrangian possesses
an additional, space dependent U(1) gauge symmetry
\be\label{spgs}
\phi(x,t)\to e^{i\theta(x)}\phi(x,t),
\ee
which indicates the absence of spatial correlations for the impurities.
This is not equivalent to the space-dependent part of the
electromagnetic U(1) gauge invariance, it only characterizes
the difference between the slowly moving and the truly quenched
impurities.

What happens in the limit $M\to\infty$? To understand this
limit better it is useful to introduce the free
diffusion constant $D=\hbar/2M$ for the impurities. 
The impurities generate an interaction on the distance scale
$\sqrt{DT}$ during the elapse of time $T$.
When this distance shrinks well below the spacelike cutoff, the
lattice spacing, the model reaches the quenched limit as far
as the observables up to time $T$ are concerned,
and the local symmetry \eq{spgs} is approximatively realized. 
The distribution of these apparently static impurities is not
uniform, it is governed by the ensemble used in the computation.
Since we shall use the canonical ensemble, our results refer to a
thermally equilibrated and after then quenched impurity distribution.

It is easy to see that the limit $M\to\infty$ enhances
the large momentum and the low frequency
contributions of the interactions. The perturbative short distance 
contributions are strengthened by the absence of the kinetic energy 
in the denominator of the propagator. This increase of the short range 
fluctuations is the impact of the static nature of the impurities 
on the spatial disorder. The slowing down of 
the impurity motion induces non-local interactions in the time rather than
the space direction \cite{effth}. The increased sensitivity at low
frequencies arises from the suppression of the hopping between
neighboring spatial locations, the reduction of the  effective spatial
dimensionality of the impurity dynamics $d_{eff}\to0$ as $M\to\infty$.
This reduces the impurity dynamics to the sum of non-interacting
Quantum Mechanical single site problems. The result of the strong
correlations generated in time is a possible non-local effect in
time within the electron sector.

The dependence of the static observables on $M/m$ is not continuous
at $M/m=\infty$, e.g. the conductivity diverges for $M/m<\infty$
because the electrons, moved by the external electric field drag the
impurities along. The conductivity is finite only in the absence of
recoil, when $M/m=\infty$.
The discontinuity at $M/m=\infty$ is reflected in the existence
of the gauge symmetry \eq{spgs} and the singularity of the Fermi surface.
In fact, the Fermi sphere for $M<\infty$ consists of the one particle
states $p^2/2M\le\mu_i$. The Fermi momentum is $p_F\approx\rho^{1/d}$ in
$d$ dimensions where $\rho$ is the particle density which yields
$\mu_i\approx\rho^{2/d}/M$. Hence the Fermi sphere becomes a highly
singular point in the momentum space as $M\to\infty$ and the
construction of the grand canonical ensemble is not obvious. The same
problem is seen by the vanishing Fermi velocity,
$v_F\approx\rho^{1/d}/M$. The ground state is highly degenerate and
admits the gauge symmetry \eq{spgs}. Due to this complication we
shall carry out the computations below for $M=\infty$ in the
canonical ensemble for the impurities.

\section{Canonical ensemble for the impurities}
Let us consider first a system of fermionic impurities without spin.
In order to avoid singularities in the static limit
we place them on a lattice with $N$ lattice points
(see Appendix \ref{lattice}) but keep
the continuous notation whenever it is not misleading.
The hamiltonian with an external source $J$ is chosen to be time dependent,
\be
H_0(t)=\int dx\left[-{\hbar^2\over2M}\phi^\dagger(x,t)\Delta\phi(x,t)
+J(x,t)\phi^\dagger(x,t)\phi(x,t)\right]
\ee
in the Heisenberg representation\footnote{
By anticipation of the perturbation expansion in the source $J$
the expressions for the free propagator will be given for $J=0$.}.
We introduce the real time expectation values of the operator $\cal O$ 
in the canonical ensemble corresponding to the density $\rho$ as
\be\label{cexpv}
\la{\cal O}\ra_\rho={1\over Z_\rho}TrP_\rho
T\left[e^{-{i\over\hbar}\int_{-T}^TdtH_0(t)}{\cal O}\right],
\ee
where
\be
Z_\rho=TrP_\rho T\left[e^{-{i\over\hbar}\int_{-T}^TdtH_0(t)}\right],
\ee
and $P_\rho$ is the projection operator onto the subspace
of $n=\rho V$ particles ($V$ denotes the volume),
\be\label{prop}
P_\rho=\int_{-\pi}^\pi{d\alpha\over2\pi}
e^{i\alpha\int dx(\phi^\dagger\phi-\rho)}
\ee
which guarantees finite density for arbitrary mass.
The finite temperature averages can be obtained by performing
a Wick rotation in the time evolution,
\be
\la {\cal O}\ra_{\rho,\beta}={1\over Z_{\rho,\beta}}TrP_\rho
T\left[e^{-\int_0^\beta dtH_0(t)}{\cal O}\right],
\ee
with
\be
Z_{\rho,\beta}=TrP_\rho T\left[e^{-\int_0^\beta dtH_0(t)}\right].
\ee
The trace is evaluated by using localized states either in momentum
or in real space for $M<\infty$ or $M=\infty$, respectively,
\be
Tr{\cal O}=\sum_{\{n\}}\la n|{\cal O}|n\ra,
\ee
where the states $|n\ra$ are the Fock states determined by
the occupation number configuration $n$,
\be\label{state}
|n\ra=\cases{\prod_p\left(a_p^\dagger\right)^{n_p}|0\ra&
for $M<\infty$,\cr
\prod_x\left(a_x^\dagger\right)^{n_x}|0\ra&for $M=\infty$,}
\ee
and
\be
[a_p,a^\dagger_q]_+=\delta_{p,q},~~~[a_x,a^\dagger_y]_+=\delta_{x,y}.
\ee
The products of creation operators are taken in an arbitrarily
chosen but fixed order in \eq{state}. We obviously have the relations
\be\label{dense}
\la a^\dagger_pa_q\ra_\rho=\hrho\delta_{p,q},~~~
\la a^\dagger_xa_y\ra_\rho=\hrho\delta_{x,y}
\ee
for a translational invariant system, where $\hrho=a^d\rho$ stands for the
dimensionless impurity density. These results allow us
to obtain the free propagator in the canonical ensemble.
The causal propagator is defined as (cf. Appendix \ref{operat})
\be
iG(x,t,x',t')=\la T[\phi(x,t)\phi^\dagger(x',t')]\ra
\ee
giving, for a fixed volume $V$,
\be
iG(x,t,x',t')={1\over V}\sum_pe^{ip(x-x')-i\epsilon_p(t-t')}
\cases{1-\hrho&for $t>t'$,\cr-\hrho&for $t<t'$,},
\ee
where $\epsilon_p$ is the single particle excitation spectrum for
$J=0$ and
\be
iG(x,t,x',t')=\cases{(1-\hrho)\delta_{x,x'}&for $t>t'$,\cr
-\hrho\delta_{x,x'}&for $t<t'$,},
\ee
for $M<\infty$ and $M=\infty$, respectively. The Fourier transformation
\bea
G(p,q,\omega)&=&\delta_{p,q}G(p,\omega)\nonu
iG(p,\omega)&=&(1-\hrho)\int_0^\infty e^{it(\omega-\epsilon_p)}dt
-\hrho\int_0^\infty e^{-it(\omega-\epsilon_p)}dt
\eea
is performed with the usual choice of the boundary condition
\be
\int_0^\infty e^{i\omega t}dt\longrightarrow
\int_0^\infty e^{i\omega t-\delta t}dt
\ee
with $\delta\to0^+$,
\bea\label{comprop}
G(p,\omega)&=&{1-\hrho\over\omega-\epsilon_p+i\delta}
+{\hrho\over\omega-\epsilon_p-i\delta}\nonu
&=&{1\over\omega-\epsilon_p+i\delta}+2\pi i\hrho\delta(\omega-\epsilon_p),\nonu
G(x,x',\omega)&=&{\delta_{x,x'}\over\omega+i\delta}
+2\pi i\hrho\delta(\omega).
\eea
The retarded Green function can be written as
\be
G_R(p,\omega)={1\over\omega-\epsilon_p+i\delta},~~~
G_R(x,x',\omega)=\delta_{x,x'}{1\over\omega+i\delta}
\ee
and the advanced one is $G_A=G_R^*$.

In order to gain an insight into the impact of the impurities
on the dynamics consider the coupled impurity-electron system
defined by the hamiltonian
\bea
H&=&\int dx\biggl[-{\hbar^2\over2m}\psi^\dagger_a(x,t)\Delta\psi_a(x,t)
+g\phi^\dagger(x,t)\phi(x,t)\psi_a^\dagger(x,t)\psi_a(x,t)\nonu
&&+{\lambda\over4}\left(\psi_a^\dagger(x,t)\psi_a(x,t)\right)^2\biggr].
\eea
Since the impurities located at different space locations
decouple it is easy to obtain the effective theory for the
electrons where the impurity degrees of freedom appear as
local variables.
Leaving the actual computation to Appendix \ref{efftha} the final
result for the expectation value of an observable $\cal O$
between the electronic states $|\Psi_{i,e}\ra$ and $|\Psi_{f,e}\ra$
in the canonical ensemble for the static impurities
can easily be obtained as a summation over the impurity
occupation numbers $n_\hx$, $\sum_xn_\hx=n=\hrho N$, $N=V/a^d$,
\be\label{sexpv}
\la\la\Psi_{f,e}|{\cal O}|\Psi_{i,e}\ra\ra=
{1\over Z_\rho}\sum_{\{n_\hx\}}\la\Psi_{f,e}|
T\left[e^{-{i\over\hbar}\int_{-T}^TdtH_{eff}(t;\{n_\hx\})}{\cal O}\right]
|\Psi_{i,e}\ra,
\ee
where
\be
Z_\rho=\sum_{\{n_\hx\}}\la\Psi_{f,e}|T\left[
e^{-{i\over\hbar}\int_{-T}^TdtH_{eff}(t;n_\hx)}\right]|\Psi_{i,e}\ra,
\ee
and
\bea
H_{eff}(t;n_\hx)&=&\int dx\biggl[
-{\hbar^2\over2m}\psi_a^\dagger(x,t)\Delta\psi_a(x,t)
+gn(x,t)\psi_a^\dagger(x,t)\psi_a(x,t)\nonu
&&+{\lambda\over4}\left(\psi_a^\dagger(x,t)\psi_a(x,t)\right)^2\biggr].
\eea
This result shows clearly the connection with model \eq{edw}, namely
the impurities are static both models but they reach equilibrium
with the electrons in our case.

\section{Auxiliary propagators}
The expressions \eq{comprop} give the correct propagator in the
canonical ensemble but are not well suited for a perturbation expansion.
In order to generate more conventional Feynman rules we look into the
detailed way the averages \eq{dense} were achieved and introduce
some auxiliary quantities. These are obtained by postponing the
integration in the projection operator \eq{prop} and by putting
the operator of the exponent in \eq{prop} into the hamiltonian.
This amounts to the usual strategy of gauge models,
treating the integral variables of the constraints in the path
integral as dynamical variables. With this in mind we introduce the
impurity hamiltonian
\be\label{aham}
H_\alpha(t)=\int dx\left[-{\hbar^2\over2M}\phi^\dagger(x,t)\Delta\phi(x,t)
+(\alpha+J(x,t))\phi^\dagger(x,t)\phi(x,t)\right],
\ee
with the source $J$ being kept vanishing in this Section
and the expectation values given by
\be
\la{\cal O}\ra_\rho={\int_{-\pi}^\pi d\alpha e^{i\alpha\rho V}
TrT\left[e^{-{i\over\hbar}\int_{-T}^TdtH_{\alpha'}(t)}{\cal O}\right]
\over\int_{-\pi}^\pi d\alpha e^{i\alpha\rho V}
TrT\left[e^{-{i\over\hbar}\int_{-T}^TdtH_{\alpha'}(t)}\right]},
\ee
where $\alpha'=\alpha/2T$.
It is furthermore useful to introduce the propagator corresponding
to a given value of $\alpha$,
\be
G_\alpha(x,t,x',t')=TrT\left[e^{-{i\over\hbar}\int_{-T}^TdtH_{\alpha'}(t)}
T[\phi(x,t)\phi^\dagger(x',t')]\right].
\ee
One can easily find its explicit form for $M=\infty$,
\be
G_\alpha(x,x',t)=\delta_{x,x'}
e^{-i\alpha't}\left(1+e^{-i\alpha}\right)^{N-1}
\cases{1&for $t>0$,\cr-e^{-i\alpha}&for $t<0$,}
\ee
since this non-normalized expectation value corresponds to the
time evolution of a state with a single filled and $N-1$ empty sites,
see Appendix \ref{operat} for the detailed derivation.
As a check we compute the impurity density,
\be
-G(x,0,x,0^+)={\int_{-\pi}^\pi d\alpha e^{in\alpha}
\left(1+e^{-i\alpha}\right)^{N-1}e^{-i\alpha}\over
a^d\int_{-\pi}^\pi d\alpha e^{in\alpha}\left(1+e^{-i\alpha}\right)^N}.
\ee
The integration over $\alpha$ selects the particle combinations
with the desired particle number,
\be
-G(x,0,x,0^+)={\pmatrix{N-1\cr n-1}\over a^d\pmatrix{N\cr n}}
={n\over V}=\rho.
\ee

When $M<\infty$ we find
\be
G_\alpha(p,t)=e^{-i(\epsilon_p+\alpha)t}
\prod_{q\not=p}\left(1+e^{-2iT(\epsilon_q+\alpha)}\right)
\cases{1&for $t>0$,\cr-e^{-2iT(\epsilon_p+\alpha)}&for $t<0$,}
\ee
where $\epsilon_p$ denotes the one-particle energy.
The finite temperature propagator reads
\bea
G(x,t,x',t')&=&{\int_{-\pi}^\pi d\alpha e^{i\alpha\rho V}
G_{\alpha/\beta}(x,t,x',t')\over
\int_{-\pi}^\pi d\alpha e^{i\alpha\rho V}
\la1\ra_{\alpha/\beta}}\nonu
&=&{1\over V}\sum_pe^{ip(x-x')-i\epsilon_p(t-t')}
{\int_{-\pi}^\pi d\alpha e^{in\alpha}
\prod_{q\not=p}\left(1+e^{-\beta\epsilon_q-i\alpha}\right)
e^{-\beta\epsilon_p-i\alpha}\over
\int_{-\pi}^\pi d\alpha e^{in\alpha}
\prod_q\left(1+e^{-\beta\epsilon_q-i\alpha}\right)}
\eea
which is just the Gibbs average of single particle contributions
in the given particle number sector.
Note that the parameter $\alpha$ is purely imaginary in an imaginary
time formalism, it appears as the time component of a gauge field.

\section{Perturbation expansion}
We have two different realizations of the canonical averages
depending on whether the projection operator is inserted
once or after each $dt$ time step\footnote{
dt serves as the ultraviolet cutoff needed in the
derivation of the path integral formulae. The limit $dt\to0$
is convergent provided that the number of the degrees of freedom
is kept finite by a cutoff in space.}
during the time evolution
$-T<t<T$. The first case leads to the path integral expression
\be\label{stat}
{1\over Z_\rho}\int d\alpha\int D[\phi]D[\phi^*]
e^{i\int dtdx[i\dot\phi^*\phi
-{\hbar^2\over2M}\phi^*\Delta\phi+\alpha(\phi^*\phi
-\rho)+J\phi^*\phi]}\cal O,
\ee
for the canonical average of the operator $T[\cal O]$, where
\be
Z_\rho=\int d\alpha\int D[\phi]D[\phi^*]
e^{i\int dtdx[i\dot\phi^*\phi
-{\hbar^2\over2M}\phi^*\Delta\phi+\alpha(\phi^*\phi-\rho)+J\phi^*\phi]}.
\ee
In the second case we have a time dependent $\alpha(t)$
trajectory to integrate over,
\be\label{nstat}
{1\over Z_\rho}\int D[\alpha]D[\phi]D[\phi^*]
e^{i\int dtdx[i\dot\phi^*\phi
-{\hbar^2\over2M}\phi^*\Delta\phi+\alpha\phi^*\phi
-\alpha\rho+J\phi^*\phi]}\cal O.
\ee
and
\be
Z_\rho=\int D[\alpha]D[\phi]D[\phi^*]
e^{i\int dtdx[i\dot\phi^*\phi
-{\hbar^2\over2M}\phi^*\Delta\phi+\alpha\phi^*\phi
-\alpha\rho+J\phi^*\phi]}.
\ee
Due to the periodicity of the integrand the
integration over $\alpha$ can be performed over the whole
real axis. The steps leading to \eq{stat} and \eq{nstat}
are similar to those giving the path integral expressions
in QED for the static temporal gauge, $\partial_0A_0=0$
and for the real temporal gauge $A_0=0$, respectively,
$A_0(x,t)$ playing the role of the projection operator
parameter $\alpha(t)$.

The straight perturbation expansion for \eq{nstat} yields
\bea
\la{\cal O}\ra&=&{1\over Z_\rho}
\int D[\alpha]D[\phi]D[\phi^*]e^{i\int dtdx[i\dot\phi^*\phi
-{\hbar^2\over2M}\phi^*\Delta\phi-\alpha\rho]}\nonu
&&\times{\cal O}\sum_n{1\over n!}\left(i\int dtdx
(\alpha+J)\phi^*\phi\right)^n,
\eea
and
\bea
Z_\rho&=&
\int D[\alpha]D[\phi]D[\phi^*]
e^{i\int dtdx[i\dot\phi^*\phi
-{\hbar^2\over2M}\phi^*\Delta\phi-\alpha\rho]}\nonu
&&\times\sum_n{1\over n!}\left(i\int dtdx(\alpha+J)\phi^*\phi
\right)^n.
\eea
The small parameter of the perturbation expansion is $J$ which
stands for
the interaction with an external source. The small parameter
for the projection operator is $1\over V$ for bosons
when the expansion is performed around $\phi(x,t)=\sqrt{\rho}$.

For fermions there is no simple way of saturating the
path integral and the projection operator must be implemented
nonperturbatively. This can be achieved when it is inserted
once only in the average,
\bea
\la{\cal O}\ra&=&{1\over Z_\rho}\int d\alpha\int D[\phi]D[\phi^*]
e^{i\int dtdx[i\dot\phi^*\phi-{\hbar^2\over2M}\phi^*\Delta\phi
+\alpha\phi^*\phi-{\alpha\over2T}\rho]}\nonu
&&\times{\cal O}\sum_n{1\over n!}\left(i\int dtdxJ\phi^*\phi\right)^n,
\eea
and
\bea
Z_\rho&=&
\int d\alpha\int D[\phi]D[\phi^*]e^{i\int dtdx[i\dot\phi^*\phi
-{\hbar^2\over2m}\phi^*\Delta\phi
+\alpha\phi^*\phi-{\alpha\over2T}\rho]}\nonu
&&\times\sum_n{1\over n!}\left(i\int dtdxJ\phi^*\phi\right)^n.
\eea

In what follows we shall use this formalism, where the hamiltonian is
given by \eq{aham}. There is a vertex
corresponding to the interaction term $J\phi^*\phi$,
the propagator is $G_\alpha$ and the integration over $\alpha$ is
to be done after the loop integrations.
The usual Feynman rules are applicable for the computation of the
integrand for the integration over $\alpha$ in the numerator and the
denominator independently. Due to the independent integrations over
$\alpha$ the disconnected diagrams do not always simplify in the
expectation values, a remnant of the non-local nature of the
canonical ensemble.
But one can verify that the grand canonical result where the
disconnected contributions simplify is recovered in the thermodynamical
limit.

A short discussion is now in order about the use of the lattice regularization
in the computation of the loop integrals emerging from the perturbation
expansion. Non-relativistic quantum field theory is non-renormalizable
in itself, a problem which is made even more serious by sending the
impurity mass to infinity as mentioned in Section \ref{model}. This
is naturally a formal problem only since the cutoff is actually kept
finite in effective theories. Nevertheless it is useful to
distinguish observables which diverge from those which stay finite
when the cutoff is removed because the computation of the
latter is simpler. To understand the reason let us start with the
remark that the lattice
regulated model is more complicated than the one obtained in
the continuum due to the trigonometric functions in the
propagator. In fact, the rule of generating the lattice propagators from
the continuum say for electrons in $2D$ is
\be
{1\over\omega-{p_2^2\over2m}+\mu+i\delta_{p_2}^\mu}\longrightarrow
{1\over\omega-{2\over ma^2_s}(\sin^2{p_xa_s\over2}+\sin^2{p_ya_s\over2})
+\mu+i\delta_{p_2}^\mu},
\ee
where $a_s$ is the spatial lattice spacing and
\be
\delta_{p_2}^\mu=\cases{0^+&${p_2^2\over2m}>\mu,$\cr
0^-&${p_2^2\over2m}<\mu$.}
\ee
In computing the integral over the spatial momentum on the lattice
one usually assumes that the cutoff is far from the internal scale
of the model, takes the limit $a_s\to0$ and expands the lattice
propagators. We regain the continuum propagator in this manner
and the higher order terms in this expansion, the irrelevant,
higher order derivative operators are formally suppressed
by a positive power of the lattice spacing. One is tempted to
ignore them altogether. But this is not always allowed because the
ultraviolet divergences of the loop integration might
generate so strong singularities that quantities which seemed to
disappear turn out to be finite or even divergent as $a_s\to0$.

One can see \cite{riesz} that no problem arises if one
considers observables in a renormalizable model without anomaly.
We can always carry out the substractions in the loop integrands
of a renormalizable theory in such a manner that the resulting
integrals are finite. An anomaly appears when "by accident" a
graph with non-negative primitive degree of divergence
given by the power counting happens to be finite and receives no
subtraction during the renormalization process. Thus in a
renormalizable, non-anomalous model all loop integrals
of the renormalized perturbation expansion are finite
and have negative primitive degree of divergence. These
integrals converge uniformly as the cutoff is removed
and the order of the integration and the limit $a_s\to0$
can be interchanged. By setting $a_s=0$ in the integrands one
eliminates all lattice artifacts and the loop integrals which
follow reproduce the continuum perturbation expansion.

Returning to our non-relativistic model, the lesson of the
argument about the suppression of the lattice artifact is that
the continuum propagators can safely be used for quantities
which stay finite as $a_s\to0$. This is enough to simplify the
computation of several important quantities such as the imaginary
part of the self energies. For other observables, the
cutoff $a_s$ must be kept finite and the ultraviolet details
of the model remain important.

\section{Electron propagation}
In this section we calculate the first two order contributions to the
electron self energy in two dimensions. The goal of this calculation
is to show that the diagrams with at most two impurity lines
atached to each vertex agree in the present approach
and those of the Edwards model can be brought into equivalence
after the rescaling of the impurity density.

We consider the case $\lambda=0$ for simplicity when the
Feynman rules can be read off from the generating functional
\bea
Z_\alpha[J^*,J,j^\dagger,j]&=&
\int D[\psi^\dagger]D[\psi]D[\phi^\dagger]D[\phi]e^{i\int dtdx
[\phi^*(i\partial_t-\alpha')\phi+\psi^\dagger_a(i\partial_t+{\hbar\over2m}\Delta
+\mu)\psi_a]}\nonu
&&\times e^{i\int dtdx[J^*\phi+\phi^*J+j^\dagger_a\psi_a+\psi_a^\dagger j_a]}
e^{-i{g\over\hbar}\int dtdx\psi^\dagger_a\psi_a\phi^*\phi}\nonu
&=&\det[(i G_\alpha)^{-1}]\det[(i G_0)^{-1}]
e^{-{ig\over\hbar}\int dtdx{\delta\over\delta J^*(x,t)}
{\delta\over\delta J(x,t)}{\delta\over\delta j^\dagger(x,t)}
{\delta\over\delta j(x,t)}}\nonu
&&\times e^{i\int dtdxdt'dx'[J^*(x,t)G_\alpha(x,t,x',t')J(x',t')
+j^\dagger(x,t)G_0(x,t,x',t')j(x',t')]}.
\eea
The electron self energy in the leading order is computed in
Appendix \ref{leadingo} and Eq. \eq{leadpr} yields
\be\label{first}
\Sigma^{(1)}=\rho g.
\ee
The two $\ord{(g^2)}$ contributions, shown in Figs. \ref{stoe}, \ref{stok} and computed
in Appendix \ref{secord} give
\be\label{sing}
Im\Sigma_1^{(2)}(\omega,p)=-{1\over\epsilon}g^2\rho(1-\hat{\rho})n_0(\mu),
\ee
and
\be
Im\Sigma_2^{(2)}(\omega,p)=-g^2\rho(1-\hat{\rho}){m\over 2}
\left(2\theta(\omega)-\theta(\omega+\mu)\right),
\ee
respectively where $\epsilon=0^+$, and
\be
n_0(\mu)={m\mu\over2\pi}.
\ee

There are peculiarities about the graph in Fig. \ref{stoe}. First,
it involves the feedback of the electron dynamics on the impurities
and it is thereby suppressed in the replica method. What is the role of
this contribution in our model? The other peculiarity
is that the meaningless $1/\epsilon$ in Eq. \eq{sing} indicates an
infrared divergence, expected to appear in zero dimensional
systems. We show that the solution of the latter problem
is the answer for the question raised before, namely it removes this
particular graph altogether from our model. The point is that
the divergence can be removed by the usual ring diagram, RPA resummation,
by taking into account the "screening" of the electron density in the
impurity propagator so that
\be\label{iprenorm}
H_\alpha(t)=\alpha'\int dx\phi^\dagger(x,t)\phi(x,t)\rightarrow
H_{\alpha+2Tgn_0(\mu)}(t)=\left(\alpha'+gn_0(\mu)\right)
\int dx\phi^\dagger(x,t)\phi(x,t).
\ee
This procedure yields
\bea
iG_{\alpha+2Tgn_0(\mu)}(x,t,x',t')
&=&\delta(x-x')e^{-i(\alpha+2Tgn_0(\mu))(t-t')}\\
&\times&\left\{\theta(t-t'-\eta)\left(1-n_{\alpha+2Tgn_0(\mu)}\right)
-\theta(t'-t+\eta)n_{\alpha+2Tgn_0(\mu)}\right\}\nonumber.
\eea
The correction to the electron propagator is therefore
\bea
&&{1\over Z_\rho}\int_{-\pi}^\pi d\alpha e^{i\alpha n}
\det[(iG_{\alpha+2Tgn_0(\mu)})^{-1}]i{g\over\hbar}\int dxiG_0(x_1,x)
iG_0(x,x_2)iG_{\alpha+2Tgn_0(\mu)}(x,x)\nonu
&=&i{g\over\hbar a_s^2}\frac{\int_{-\pi}^\pi d\alpha
e^{i\alpha n}(\left[1+e^{-i(\alpha+2Tgn_0(\mu))})\right]^{N-1}
e^{-i(\alpha+2Tgn_0(\mu))}}{\int_{-\pi}^\pi d\alpha
e^{i\alpha N}\left[1+e^{-i(\alpha+2Tgn_0(\mu))}\right]^N}
\int dxG_0(x_1,x)G_0(x,x_2)\nonu
&=&i{g\over\hbar a_s^2}{\pmatrix{N-1\cr n-1}\over\pmatrix{N\cr n}}
\int dxG_0(x_1,x)G_0(x,x_2)\nonu
&=&i{\rho g\over\hbar}\int dxG_0(x_1,x)G_0(x,x_2).
\eea
The result of such a partial resummation is finite and agrees with
\eq{first}. This is because $G_\alpha$ is a periodic function
of $\alpha$ and the integration, carried out over the length of a
period, is invariant under the shift $\alpha\to\alpha+c$. Thus
the only effect of the renormalization \eq{iprenorm} is the
omission of the contribution of the graph of
Fig. \ref{stoe} to the RPA ring diagrams altogether. In other words,
the discretness of the electron number removes the feedback of
the electron dynamics on the impurities.

The complete $\ord{(g^2)}$ self energy after resummation is
\be\label{imsi}
Im\Sigma^{(2)}(\omega)=-g^2\rho(1-\hat{\rho}){m\over 2}
\left(2\theta(\omega)-\theta(\omega+\mu)\right),
\ee
which is independent of the momentum $p$ because the impurity dynamics
is ultralocal, the pointlike impurities do not propagate.
The imaginary part of the self energy is related to the mean free path,
in particular for the electrons above of the Fermi level, $\omega>0$
we have
\be
Im\Sigma(\omega)=-g^2\rho(1-\hat\rho){m\over2}+\ord{(g^3)}
\approx-{\hbar\over2\tau},
\ee
where $\tau$ is the relaxation time. Recall that the dimensionless
density $\hat\rho\in[0,1]$ and $Im\Sigma=\ord{(\rho)}$, the electron
relaxation time diverges at low impurity densities.
The analogous expression of the Edwards model can be obtained
from \eq{imsi} by the replacement
\be\label{resc}
\rho(1-\hrho)\longleftrightarrow\rho.
\ee
Our expression reflects the fact that the electrons do not scatter on
the completely filled up impurity system (Pauli blocking), a phenomenon
which is neglected when the effects of the impurities is represented
by a static potential only.

Let us now consider the propagation of a particle-hole pair with
energy-momenta $\omega_+,p_+$, $\omega_-,p_-$, respectively, and write
the corresponding amplitude as
\bea
G(\omega_+,p_+,\omega_-,p_-,\omega'_+,p'_+,\omega'_-,p'_-)
&=&iG_0(\omega_+,p_+)iG_0(\omega'_+,p'_+)iG_0(\omega_-,p_-)
iG_0(\omega'_-,p'_-)\nonu
&&\times\delta(p_+-p'_+-p_-+p'_-)\Xi.
\eea
In the leading order, before the $\alpha$ integration we find
\bea
\Xi_\alpha&=&(ig)^2\sum_{\omega,\omega'}iG_\alpha(\omega)iG_\alpha(\omega')
\delta(\omega_+-\omega'_++\omega-\omega')
\delta(\omega'-\omega+\omega'_-\omega_-)\nonu
&=&-(ig)^22\pi(1-\hrho_\alpha)\rho_\alpha.
\eea
The $\alpha$ integration yields
\be
\Xi=-(ig)^22\pi(1-\hrho)\rho,
\ee
a result which agrees with the Edwards model after the
replacement \eq{resc}. The insertion of $\ell$ impurity
bubbles at different sites and the $\alpha$ integration yield the factor
\be\label{ninsert}
\frac{\int_{-\pi}^\pi e^{i\alpha n}\det[(iG_\alpha)^{-1}]
\left[(1-n_\alpha)n_\alpha\right]^\ell}
{\int_{-\pi}^\pi e^{i\alpha n}\det[(iG_\alpha)^{-1}]}
\approx\left[\rho(1-\hrho)\right]^\ell.
\ee
Since the particle-hole lines are the usual ones in the rest of the
diagram, this leads to the equivalence of the resummation over the
ladder or the maximally crossed graphs, as well.
In the light of the results of ref. \cite{vowo} this
indicates that the localisation-delocalisation phase structure is
similar in these descriptions.

Not all diagrams of the local, Gaussian potential model
\eq{edw} can be brought into equivalence with our model.
There one finds contributions with more than two impurity lines
at a coordinate space location. This can not happen in our model
where the fermionic statistic of the impurity cancels these diagrams.
But it is easy to see that the remaining diagrams,
where only two impurity lines are attached at each
spatial lattice site agree after the change \eq{resc}.
Furthermore, the disconnected contributions simplify for
these graphs in the thermodynamic limit. The point is that
the impurity propagators factorise and the disconnected parts drop out
as in the usual, annealed averages. The resulting impurity propagators
at different lattice sites give the same contributions when the
thermodynamic limit is performed. This result is actually expected
if the canonical and the grand canonical ensemble are equivalent.

There are furthermore diagrams in our model which have no counterpart
in \eq{edw}. These graph contain internal fermion loops and
some of them cancel after resummation. But there are remaining
non-vanishing contributions, e.g. the
correlation of the potential is ultralocal and
unchanged by the electrons for \eq{edw} but the electron dynamics
generates a non-trivial correlation function for the impurity
density ($\phi^*\phi$) in our model, the remnant of the annealed
integration over the field $\phi$. Such an electron dynamics
generated correlation is absent in the Green functions for the
impurities, the correlation functions of the field variable
$\phi^*$ or $\phi$ remain local due to the gauge symmetry \eq{spgs}.
It remains to be seen if the feedback of the electron dynamics
generated by the resummation of the irreducible vertices
obtained in the higher loop aproximation cancels or not.

\section{Conclusions}
The quenched averaging over static impurities is achieved
within the second quantized formalism by means of an annealed
averaging procedure where the impurity motion is slowed down by setting
the impurity mass equal to infinity. This description corresponds to
a system of electrons and static impurities in equilibrium.
Despite the feedback of the electrons on the impurity dynamics
the localisation-delocalisation phase structure in the present approach 
is expected to be similar to the phase structure of the Edwards model.

The problem of the usual methods for disordered systems is
their inability to deal with strong interactions. However our model
is well suited to non-perturbative methods. The numerical simulation
in lattice regularization is feasible since
one can easily construct stochastic sampling algorithms in our
canonical ensemble at finite temperature. Another natural
non-perturbative method available is the functional
renormalization group approach in the internal space \cite{int}
which can treat the finite density fermionic systems in a simple
manner.

\section{Acknowledgment}
We thank A. Patk\'os and F. Wegner for useful discussions.
The work was supported in part by the grant OTKA T29927/98
of the Hungarian Academy of Sciences.

\begin{appendix}
\section{Lattice regularization}\label{lattice}
The regulator applied in this work is a lattice in space-time with
spatial and temporal lattice spacing $a_s$, and $a_t$, respectively.
The action is written as
\be
S={1\over\hbar}\int dtdx
L[\phi^\dagger(x,t),\phi(x,t),\psi^\dagger(x,t),\psi(x,t)]\longrightarrow
\sum_{\hx,\hti}\hat L[\hphi^\dagger_{\hx,\hti},\hat\phi_{\hx,\hti},
\hat\psi^\dagger_{\hx,\hti},\hat\psi_{\hx,\hti}],
\ee
with $S=S_i+S_e+S'$,
\bea
\hbar\int dtdx\phi^\dagger(x,t)(i\partial_t-\alpha)\phi(x,t)&\to&
\hat{S}_i=\sum_{\hx,\hti}\hphi_{\hx,\hti}^\dagger
(i\hat\partial_t-\hat\alpha)\hat{\phi}_{\hx,\hti},\nonu
\int dtdx\psi^\dagger(x,t)(i\partial_t+{\hbar\over2m}\Delta+\mu)\psi(x,t)&\to&
\hat S_e=\sum_{\hx,\hti}\hpsi_{\hx,\hti}^\dagger
(i\hat\partial_t+{\hat\hbar\over2\hat m}\hat\Delta+\hat\mu)
\hpsi_{\hx,\hti},\nonu
-g\hbar\int dtdx\psi^\dagger(x,t)\psi(x,t)\phi^\dagger(x,t)\phi(x,t)&\to&
\hat S'=-\hat g\sum_{\hx,\hti}\hpsi_{\hx,\hti}^\dagger
\hpsi_{\hx,\hti}\hphi_{\hx,\hti}^\dagger
\hphi_{\hx,\hti},
\eea
where $\hx^j=n^ja_s$, $j=1,\cdots,d$, $\hti=\ell a_t$ and
the following lattice quantities were introduced in $2D$
$\hphi_{\hx,\hti}=a_s\phi(\hx,\hti)$, (the field variables
with space-time coordinates in the subscript are dimensionless)
$\hat\alpha=a_t\alpha$, $\hat m=a_sm$, $\hat\hbar=a_t\hbar/a_s$,
$\hpsi_{\hx,\hti}=a_s\psi(\hx,\hti)$, $\hat g=g/a_s$,
and $\hat\mu=a_t\mu$. The finite difference operators are defined as
$\hat\partial_\tau f_{n,\ell}=f_{n,\ell}-f_{n,\ell-1}$ and
$\hat\Delta f_{n,\ell}=\sum_{i=1}^df_{n+e_i,\ell}+f_{n-e_i,\ell}-2f_{n,\ell}$.

\section{Operator formalism}\label{operat}
The dynamics generated by the free impurity hamiltonian
\be\label{frham}
H_\alpha=\hat\alpha\sum_\hx\hat n_\hx,
\ee
where the static impurity density is denoted by
\be
\hat n_\hx=\hphi^\dagger_{\hx,t}\hphi_{\hx,t}
\ee
is simplest to discuss in the operator formalism.
The impurity propagator is written as
\be
iG_\alpha(x,t,x',t')={1\over a_s^2}\hat G_{\alpha,\hx,\hti,\hx',\hti'}
={1\over\cal N}\sum_{\{n\}}
\la n|T[e^{-{i\over\hbar}\int_{-T}^TdtH_{\alpha'}(t)}
\hphi(\hx,t)\hphi^\dagger(\hx',t')]|n\ra,
\ee
where
\be
|n\ra=\prod_\hx\left(\hphi_\hx^\dagger\right)^{\hat n_{\hx}}|0\ra.
\ee
and ${\cal N}$ is a normalization constant. The Hamiltonian \eq{frham} gives
\bea
i{\cal N}\hat G_{\alpha,\hx,\hti,\hx',\hti'}&=&
\sum_{\{n\}}\theta(t-t')\la n|e^{-i\alpha'\hat{n}(T-t)}\hphi_\hx
e^{-i\alpha'(t-t')}\hphi^\dagger_{\hx'}e^{-{i\alpha}(t'+T)}|n\ra\nonu
&&-\sum_{\{n\}}\theta(t'-t)\la n|e^{-i\alpha'(T-t')}\hphi^\dagger_{\hx'}
e^{-i\alpha'(t'-t)}\hphi_\hx e^{-i\alpha'(t+T)}|n\ra.
\eea
Since the lattice sites are decoupled we find immediately
\bea
&\sum_{\{n\}}\la n|e^{-i\alpha'(T-t)}\hphi_\hx
e^{-i\alpha'(t-t')}\hphi^\dagger_{\hx'}e^{-i\alpha'(t'+T)}|n\ra
&=e^{-i\alpha'(t-t')}\delta_{\hx,\hx'}(1+e^{-i\alpha})^{N-1},\nonu
&\sum_{\{n\}}\la n|e^{-i\alpha'(T-t')}\hphi^\dagger_{\hx'}
e^{-i\alpha'(t'-t)}\hphi_\hx e^{-i\alpha'(t+T)}|n\ra
&=e^{-i\alpha'(t-t')}\delta_{\hx,\hx'}(1+e^{-i\alpha})^{N-1}e^{-i\alpha T},
\eea
where $N$ denotes the number of sites. The normalization constant is
\be
{\cal N}=\det[iG_\alpha]^{-1}=(1+e^{-i\alpha})^N,
\ee
yielding
\be\label{lpropa}
i\hat G_{\alpha,\hx,\hti,\hx',\hti'}
=e^{-i\alpha'(t-t')}\delta_{\hx,\hx'}
(1+e^{-i\alpha})^{-1}\left(\theta(t-t')-\theta(t'-t)e^{-i\alpha}\right)
\ee
on the lattice or
\be
iG_\alpha(x,t,x',t')=\delta(x-x')e^{-i\alpha'(t-t')}
\left(\theta(t-t')(1-n_\alpha)-\theta(t'-t)e^{-i\alpha}n_\alpha\right)
\ee
in the continuum where
\be
n_\alpha={1\over1+e^{i\alpha}}.
\ee

\section{Functional approach}\label{funct}
The interactions are easier to take into account in the path integral
representation. To make sure that all singular features of the
limit of diverging impurity mass are properly kept we first
quickly reproduce the free impurity propagator in this representation.
The amplitude
\be\label{freez}
Z_{0\alpha}=Tre^{-i\int_{-T}^TdtH_{\alpha'}(t)}
\ee
can be written as a functional integral over Grassmannian coherent state
configurations which are antiperiodic in time by inserting the resolution
of the identity
\be
1=\int\prod_x d\phi^*(x,t)d\phi(x,t)e^{-\int dx\phi^*(x,t)\phi(x,t)}
|\phi(x,t\ra\la\phi(x,t)|,
\ee
into \eq{freez} at different times $t_m$,
\bea
Z_{0\alpha}&=&\lim_{N_t\to\infty}\int\prod_{m=-N_t+1}^{N_t}
\prod_xd\phi^*(x,t_m)
d\phi(x,t_m)e^{-\int dx\phi^*(x,t_m)\phi(x,t_m)}\nonu
&&\times\prod_{m=-N_t+1}^{N_t}\la\phi(x,t_m)|e^{-ia_tH_{\alpha'}(t)}|\phi(x,t_m)\ra\nonu
&=&\lim_{N_t\to\infty}\int\prod_{m=-N_t+1}^{N_t}
\prod_xd\phi^*(x,t_m)d\phi(x,t_m)
e^{-\sum_{m=-N_t+1}^{N_t}\int dx\phi^*(x,t_m)\phi(x,t_m)}\nonu
&&\times e^{-\sum_{m=-N_t+2}^{N_t}
\int dx\phi^*(x,t_m)(1-ia_t\alpha')\phi(x,t_{m-1})}
e^{-\int dx\phi^*(x,t_{-N_t+1})(1-ia_t\alpha')\phi(x,t_{N_t})}\nonu
&=&\prod_x\bigl[\lim_{N_t\to\infty}\int\prod_{m=-N_t+1}^{N_t}
d\phi^*(x,t_m)d\phi(x,t_m)\nonu
&&\times e^{-\sum_{m,n=-N_t+1}^{N_t}\phi^*(x,t_m)S_{mn}\phi(x,t_n)}\bigr],
\eea
where the matrix $S$ is
\be
S=\pmatrix{1 &0  &\cdots  &\cdots &\cdots &0      &A\cr
	-A&1  &0       &\cdots &\cdots &\cdots &0\cr
	0 &-A &1       &0      &\cdots &\cdots &0\cr
	\cdots&\cdots&	\cdots	&\cdots	&\cdots	&\cdots	&\cdots\cr
	\cdots&\cdots&	\cdots	&\cdots&\cdots	&\cdots	&\cdots\cr
	\cdots&\cdots&	\cdots	&\cdots	&\cdots	&\cdots	&\cdots\cr
	\cdots&\cdots&	\cdots	&\cdots	&\cdots	&-A	&1},
\ee
with $A=1-ia_t\alpha'=1-i\alpha/2N_t$. The Grassmann integration gives
\be
Z_{0\alpha}=\prod_x\left[\lim_{N_t\to\infty}\det S\right]
=\prod_x\lim_{N_t\to\infty}\left[1+A^{2N_t}\right]
=\left(1+e^{-i\alpha}\right)^N.
\ee
The propagator is given for $t_m>t_{m'}$ by
\bea
i\hat G_{\alpha,\hx,\hx'}(t_m,t_{m'})
&=&\delta_{\hx,\hx'}\lim_{N\to\infty}\left(S^{-1}\right)_{mm'}\nonu
&=&\delta_{\hx,\hx'}\lim_{N\to\infty}\frac{A^{m-m'}}{1+A^{2N_t}}\nonu
&=&\delta_{\hx,\hx'}e^{-i\alpha(t_m-t_{m'})}
\left(1-{1\over1+e^{i\alpha}}\right),
\eea
and for $t_m<t_{m'}$ by
\bea
iG_{\alpha,\hx,\hx'}(t_m,t_{m'})
&=&\delta_{\hx,\hx'}\lim_{N_t\to\infty}\left(S^{-1}\right)_{mm'}\nonu
&=&\delta_{\hx,\hx'}\lim_{N_t\to\infty}\frac{-A^{2N_t+m-m'}}{1+A^{2N_t}}\nonu
&=&-\delta_{\hx,\hx'}e^{-i\alpha(t_m-t_{m'})}{1\over1+e^{i\alpha}},
\eea
in agreement with \eq{lpropa}.

The Fourier transform of the propagator in time is
\bea
G_\alpha(x,x',\omega)&=&\int d(t-t')e^{i\omega(t-t')}G_\alpha(x,t,x',t')\nonu
&=&\delta(x-x')\lim_{\eta\to0^+}e^{i\omega\eta}\left(\frac{1-n_\alpha}
{\omega-\alpha'+i\epsilon}+\frac{n_\alpha}{\omega-\alpha'-i\epsilon}\right)
\eea
in the continuum and
\bea
\hat G_{\alpha,\hx,\hx'}(\omega_m)
&=&\delta_{\hx,\hx'}\lim_{\eta\to0^+}e^{i\omega_m\eta}\left(
\frac{1-n_\alpha}{{\pi m\over T}-\hat\alpha'+i\epsilon}+\frac{n_\alpha}
{{\pi m\over T}-\hat\alpha'-i\epsilon}\right)
\eea
on the lattice where the infinitesimal quantity $\eta$ was introduced to
take care about the point splitting.

We shall need later
\bea
iG_\alpha(x,0;x,0^+)
&=&\lim_{\eta\to 0^+}\int_{-\infty}^\infty{d\omega\over2\pi}\,
iG_\alpha(x,x;\omega)e^{i\omega\eta}\nonu
&=&{i\over a_s^2}\lim_{\eta\to 0^+}\int_{-\infty}^\infty{d\omega\over2\pi}\,
\left(\frac{1-n_\alpha}{\omega-\alpha'+i\epsilon}+
\frac{n_\alpha}{\omega-\alpha'-i\epsilon}\right)e^{i\omega\eta}\nonu
&=&{i\over a_s^2}\lim_{\eta\to0^+}\int_{-\infty}^\infty{d\omega\over2\pi}\,
\frac{n_\alpha e^{i\omega\eta}}{\omega-\alpha'-i\epsilon}\nonu
&=&-{1\over a_s^2}(1+e^{-{i\alpha}})^{-1}e^{-i\alpha}
\eea
which, after integration over $\alpha$ leads to
\bea
-iG(x,0,x,0^+)&=&{1\over a_s^2}{\int_{-\pi}^\pi d\alpha e^{in\alpha}
\left(1+e^{-i\alpha}\right)^{N-1}e^{-i\alpha}\over
\int_{-\pi}^\pi d\alpha e^{in\alpha}\left(1+e^{-i\alpha}\right)^N}\nonu
&=&\frac{n}{Na_s^2}\nonu
&=&\rho.
\eea

\section{Effective theory}\label{efftha}
Consider now the transition amplitudes of the
impurity-electron coupled system in the path integral representation
\bea\label{lpint}
\la\Psi_{f,e},\Psi_{f,i}|e^{-i{t\over\hbar}H}|\Psi_{i,e},\Psi_{i,i}\ra&=&
\prod_{\hx,\hti}\int d\hpsi^*_{\hx,\hti,a}
d\hpsi_{\hx,\hti,a}d\hphi^*_{\hx,\hti}d\hphi_{\hx,\hti}\nonu
&&\Psi^*_{f,e}[\hpsi^*,\hpsi]\Psi^*_{f,i}[\hphi^*,\hphi]
\Psi_{i,e}[\hpsi^*,\hpsi]\Psi_{i,i}[\hphi^*,\hphi]
e^{-{i\over\hbar}S[\hpsi^*,\hpsi,\hphi^*,\hphi]},
\eea
of the model \eq{mham} on a discretised space-time lattice, where
\bea\label{mlagl}
{1\over\hbar}S[\hpsi^*,\hpsi,\hphi^*,\hphi]&=&
\sum_{\hx,\hti}\biggl[i\hpsi^*_{\hx,\hti,a}
(\hpsi_{\hx,\hti,a}-\hpsi_{x\hat ,t-1,a})\nonu
&&+{\hat\hbar\over2\hat m}\sum_{\hat i}
\hpsi^*_{\hx,\hti,a}(\hpsi_{\hx+a_s\hat i,\hti-a_t,a}
+\hpsi_{\hx-a_s\hat i,\hti-a_t,a}-2\hpsi_{\hx,\hti-a_t,a})\\
&&-{\lambda\over4}(\hpsi^*_{\hx,\hti,a}
\hpsi_{\hx,\hti-a_t,a})^2\nonu
&&+i\hphi^*_{\hx,\hti}(\hphi_{\hx,\hti}-\hphi_{\hx,\hti-a_t})
+\hat g\hpsi^*_{\hx,\hti,a}\hpsi_{\hx,\hti-a_t,a}
\hphi^*_{\hx,\hti}\hphi_{\hx,\hti-a_t}\biggr].\nonumber
\eea
The wave functionals of the initial and the final states are
$\Psi_{i,e}[\hpsi^*,\hpsi]$, $\Psi_{i,i}[\hphi^*,\hphi]$,
$\Psi^*_{f,e}[\hpsi^*,\hpsi]$, and $\Psi^*_{f,i}[\hphi^*,\hphi]$.

Notice that the impurity field is coupled in the time direction only and
the functional integral over $\phi^*$ and $\phi$ decouples into a single
site, quantum mechanical problem. This suggests the introduction
of an effective theory obtained after integration over the impurity
field,
\be\label{effpi}
e^{-{i\over\hbar}S_{eff}[\hpsi^*,\hpsi]}=
\prod_{\hx,\hti}\int d\hphi^*_{\hx,\hti}d\hphi_{\hx,\hti}
\Psi^*_{f,i}[\hphi^*,\hphi]\Psi_{i,i}[\hphi^*,\hphi]
e^{-{i\over\hbar}S[\hpsi^*,\hpsi,\hphi^*,\hphi]}.
\ee
This is the path integral of a two level system with hamiltonian
$H_t=J_t$, where
\be
J_t=\hat g\hpsi^*_{\hx,\hti,a}\hpsi_{\hx,\hti-a_t,a}.
\ee
The matrix elements
\bea
\la0|e^{-i{t\over\hbar}H}|0\ra&=&1,\nonu
\la1|e^{-i{t\over\hbar}H}|1\ra&=&T\left[\prod_te^{-iJ_t}\right]
\eea
agree with the path integral
\be\label{sspi}
\la n|e^{-i{t\over\hbar}H}|n\ra=
(\hphi_f^*\hphi_i)^n\prod_{\hti}\int d\hphi^*_{\hti}d\hphi_{\hti}
e^{-i[i\hphi^*_{\hti}(\hphi_{\hti}-\hphi_{\hti-a_t})
+J_{\hti}\hphi^*_{\hti}\hphi_{\hti-a_t}]},
\ee
where $\phi_i$ and $\phi_f$ stand for the first and the last
integration variable in time, the arguments of the initial and
final single site wave function
\be
\Psi_{i,i}[\hphi^*,\hphi]=\Psi_{f,i}[\hphi^*,\hphi]
=\hphi^{*n}.
\ee
In fact, \eq{sspi}
can be checked by either direct computation of the fermion
determinant or by expanding the integrand in the quasi-local term
$\phi^*_t\phi_{t-1}$. One finds the effective action
\bea\label{effae}
{1\over\hbar}S_{eff}[\hpsi^*,\hpsi]&=&
\sum_{\hx,\hti}\biggl[i\hpsi^*_{\hx,\hti,a}
(\hpsi_{\hx,\hti,a}-\hpsi_{\hx,\hti-a_t,a})\nonu
&&+{\hat\hbar\over2\hat m}\sum_{\hat i}
\hpsi^*_{\hx,\hti,a}(\hpsi_{\hx+a_s\hat i,\hti-a_t,a}
+\hpsi_{\hx-a_s\hat i,\hti-a_t,a}-2\hpsi_{\hx,\hti-a_t,a})\nonu
&&-{\lambda\over4}(\hpsi^*_{\hx,\hti,a}\hpsi_{\hx,\hti-a_t,a})^2
+\hat gn_\hx\hpsi^*_{\hx,\hti,a}\hpsi_{\hx,\hti-a_t,a}\biggr]
\eea
for the states
\be
\Psi_{i,i}[\hphi^*,\hphi]=\Psi_{f,i}[\hphi^*,\hphi]
=\prod_\hx\hphi_\hx^{*n_\hx},
\ee
which belong to the occupation number configuration $\{n_x\}$, $n_\hx=0,1$.
Notice that the impurity induced interaction is as long range in time
as possible, i.e. time independent, and ultralocal in space due to the
conservation laws originating from the symmetry \eq{spgs}.

The time ordered expectation value of an observable $\cal O$ of the electrons
can easily be obtained as a summation over the impurity
occupational number configurations $n_\hx$,
$\sum_xn_\hx=\rho N$,
\bea
\la\la\Psi_{f,e}|T[{\cal O}]|\Psi_{i,e}\ra\ra&=&{1\over Z_\rho}
\sum_{\{n_\hx\}}
\prod_{\hx,\hti}\int d\hpsi^*_{\hx,\hti,a}
d\hpsi_{\hx,\hti,a}\nonu
&&\Psi^*_{f,e}[\hpsi^*,\hpsi]\Psi_{i,e}[\hpsi^*,\hpsi]
e^{-{i\over\hbar}S_{eff}[\hpsi^*,\hpsi,n_\hx]}{\cal O},
\eea
where
\be
Z_\rho=\sum_{\{n_\hx\}}
\prod_{\hx,\hti}\int d\hpsi^*_{\hx,\hti,a}
d\hpsi_{\hx,\hti,a}
\Psi^*_{f,e}[\hpsi^*,\hpsi]\Psi_{i,e}[\hpsi^*,\hpsi]
e^{-{i\over\hbar}S_{eff}[\hpsi^*,\hpsi,n_\hx]}.
\ee

\section{Electron self energy}
\subsection{$\ord{(g)}$}\label{leadingo}
The $\alpha$-dependent electron propagator can be written in terms of
the free electron and impurity propagators $G_0$ and $G_\alpha$,
respectively as 
\bea\label{elpropa}
\la\psi(x_1)\psi^\dagger (x_2)\ra_\alpha
&=&\det[(iG_\alpha)^{-1}]\det[(iG_0)^{-1}]\nonu
&&\times\Bigg\{iG_0(x_1,x_2)\Big[1-i{g\over\hbar}
\int dxiG_\alpha(x,x)iG_0(x,x)\nonu
&&+\left(-i{g\over\hbar}\right)^2
\int dx\int dy\Big(iG_0(x,x)iG_0(y,y)i G_\alpha(x,x)iG_\alpha(y,y)\nonu
&&-iG_\alpha(x,x)iG_\alpha(y,y)iG_0(x,y)iG_0(y,x)\nonu
&&-iG_\alpha(x,y)iG_\alpha(y,x)iG_0(x,x)iG_0(y,y)\Big)\Big]\nonu
&&+i{g\over\hbar}\int dxiG_0(x_1,x)iG_0(x,x_2)iG_\alpha(x,x)\nonu
&&\times\Big[1-i{g\over\hbar}\int dyiG_0(y,y)iG_\alpha(y,y)\Big]\nonu
&&+\left(i{g\over\hbar}\right)^2
\int dx\int dy\Big(iG_0(x_1,x)iG_0(x,x_2)
iG_\alpha(x,y)iG_\alpha(y,x)iG_0(y,y)\nonu
&&-iG_0(x_1,x)iG_0(x,y)iG_0(y,x_2)iG_\alpha(x,y)iG_\alpha(y,x)\Big)\Bigg\}
+\ord{(g^3)},
\eea
where the compact notation $(x,t)\to x$ was introduced. For the normalization
we need
\bea
Z_\alpha&=&\int D[\psi^\dagger]D[\psi]D[\phi^*]D[\phi]e^{{i\over\hbar}
\int dxdt[\phi^\dagger(i\hbar\partial_t -\hbar\alpha')\phi
+\psi^\dagger(i\hbar\partial_t+{\hbar^2\over2m}\Delta+\mu)\psi]}\nonu
&&\times\Bigg(1-i{g\over\hbar}\int dtdx
\psi^\dagger(x,t)\psi(x,t)\phi^*(x,t)\phi(x,t)+\ord{(g^2)}\Bigg)\nonu
&=&\det[iG_0]^{-1}\left(1+e^{-i\alpha}\right)^N\nonu
&&\times\Bigg\{1-i{g\over\hbar}\int dxdtiG_0(xt;xt)
iG_\alpha(xt;xt)+\ord{(g^2)}\Bigg\}.
\eea
The complete propagator
\be
\la\psi(x_1)\psi^\dagger(x_2)\ra=\frac{\int_{-\pi}^\pi d\alpha e^{i\alpha n}
\la\psi(x_1)\psi^\dagger(x_2)\ra_\alpha}
{\int_{-\pi}^\pi d\alpha e^{i\alpha n}Z_\alpha}.
\ee
starts as $iG_0(x_1,x_2)$ in $\ord{(g^0)}$. The non-trivial $\ord{(g)}$ 
piece in the numerator is
\bea
&&\int_{-\pi}^\pi d\alpha e^{i\alpha n}\det[(iG_\alpha)^{-1}]
i{g\over\hbar}\int dxiG_0(x_1,x)iG_0(x,x_2)iG_\alpha(x,x)\nonu
&&\hspace{.2cm}\times\Big[1-i{g\over\hbar}\int dyiG_0(y,y)
iG_\alpha(y,y)\Big]\nonu
&&=i{g\over\hbar a^2_s}\left[\pmatrix{N-1\cr n-1}
+i{g\over\hbar a^2_s}Tr[iG_0]\pmatrix{N-2\cr n-2}\right]
\int dxG_0(x_1,x)G_0(x,x_2).
\eea
Taking into account the normalization we find
\bea\label{leadpr}
\la\psi(x_1)\psi^\dagger(x_2)\ra&=&iG_0(x_1;x_2)\nonu
&&+i{g\over\hbar a^2_s}\int dxG_0(x_1,x)G_0(x,x_2)
\frac{\pmatrix{N-1\cr n-1}+i{g\over\hbar a^2_s}Tr[iG_0]\pmatrix{N-2\cr n-2}}
{\pmatrix{N\cr n}+i{g\over\hbar a^2_s}Tr[iG_0]\pmatrix{N-1\cr n-1}}
+\ord{(g^2)}\nonu
&=&iG_0(x_1;x_2)+i{g\over\hbar}\rho\int dxG_0(x_1;x)G_0(x;x_2)+\ord{(g^2)}
\eea
in the thermodynamic limit, ($N, n\to\infty, \, n/N\to\hrho$).
The comparison with the  Schwinger-Dyson equation equation 
gives the self energy
\be
\Sigma^{(1)}=\rho g,
\ee
a simple shift in the energy. In general, the diagram with $\ell$ impurity
bubbles produces the factor 
\be
{\pmatrix{N-\ell\cr n-\ell}\over\pmatrix{N\cr n}}
={N!(n-\ell)!\over n!(N-\ell)!}\to\hat{\rho}^\ell.
\ee

\subsection{$\ord{(g^2)}$}\label{secord}
\begin{figure}
\epsfxsize=10cm 
      \centerline{\epsffile{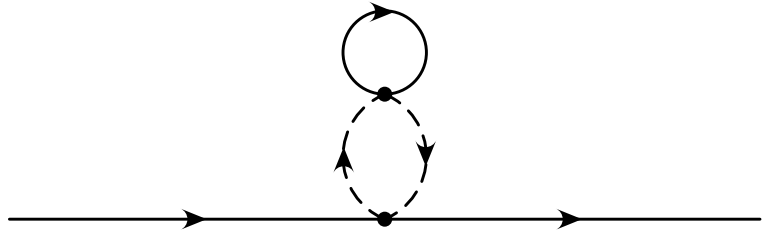}}
\caption{\label{stoe} Energy independent $\ord{(g^2)}$ 
contribution to the electron self energy.}
\end{figure}

The graph shown in Fig.\ref{stoe} is
\bea
I_{\alpha 1}&=&\int dxdtdydt'iG_0(x_1,t_1;x,t)iG_0(x,t;x_2,t_2)iG_0(y,t';y,t')
iG_\alpha(x,t;y,t')iG_\alpha(y,t';x,t)\nonu
&=&{1\over a^2_s}\int dx\int dt\int dt'iG_0(x_1,t_1;x,t)
iG_0(x,t;x_2,t_2)iG_0(y,t';y,t')\nonu
&&\times\left[\theta(t-t')\left(1-n_\alpha\right)-\theta(t'-t)n_\alpha \right]
\left[\theta(t'-t)\left(1-n_\alpha\right)-\theta(t-t')n_\alpha \right]\nonu
&=&{i^5\over a^2_s}\int{d\omega_1\over2\pi}\int{d\omega_2\over 2\pi}
\int{d\omega_3\over2\pi}\int{d^2p_1\over(2\pi)^2}\int{d^2p_2\over(2\pi)^2}
\frac{e^{ip_1(x_1-x_2)-i\omega_1(t_1-t_2)}}{(\omega_1-{p_1^2\over2m}
+\mu+i\delta_{p_1}^\mu)^2}\nonu
&&\times\frac{1}{\omega_2-{p_2^2\over2m}+\mu+i\delta_{p_2}^\mu} 
e^{2i\omega_3\eta}\left(\frac{1-n_\alpha}{\omega_3+i\epsilon}+\frac{n_\alpha}
{\omega_3-i\epsilon}\right)^2.
\eea
The integration over $\omega_3$ is divergent
\be
\int{d\omega_3\over2\pi}e^{2i\omega_3\eta}\left(\frac{1-n_\alpha}{\omega_3
+i\epsilon}+\frac{n_\alpha}{\omega_3-i\epsilon}\right)^2
={(1-n_\alpha)n_\alpha\over\epsilon},
\ee
where $\epsilon\to0^+$. The projection onto the subspace with $n$
impurities is obtained by the integration
\bea
\frac{\int_{-\pi}^\pi e^{i\alpha n}\det[(iG_\alpha)^{-1}]
(1-n_\alpha)n_\alpha}{\int_{-\pi}^\pi e^{i\alpha n}\det[(iG_\alpha)^{-1}]}
&=&\frac{\int_{-\pi}^\pi e^{i\alpha(n-1)}(1+e^{-i\alpha})^{N-2}}
{\int_{-\pi}^\pi e^{i\alpha n}(1+e^{-i\alpha})^N}\nonu
&=&\frac{\pmatrix{N-2\cr n-1}}{\pmatrix{N\cr n}}\nonu
&\approx&\hat{\rho}(1-\hat{\rho}).
\eea

The integration over $\omega_2$ and $p_2$ gives
\bea
\int{d^2p_2\over(2\pi)^2}\int{d\omega_2\over2\pi}\frac{1}{\omega_2-{p_2^2\over
2m}+\mu+i\delta_{p_2}^\mu}
&=&\int{d^2p_2\over(2\pi)^2}\lim_{t\to 0^-}\int{d\omega_2\over2\pi}
\frac{e^{-i\omega_2t}}{\omega_2-{p_2^2\over2m}+\mu+i\delta_{p_2}^\mu}\nonu
&=&\int{d^2p_2\over(2\pi)^2} i\theta(\mu-{p_2^2\over 2m})\nonu
&=&in_0(\mu).
\eea
The corresponding contribution to the self-energy, obtained after
removing the external legs is
\be\label{divep}
\Sigma_1^{(2)}(\omega,p)=-{i\over\epsilon}g^2\rho(1-\hat{\rho})n_0(\mu).
\ee

\begin{figure}
\epsfxsize=8cm 
      \centerline{\epsffile{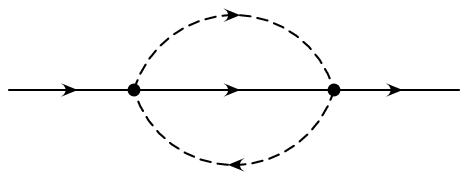}}
\caption{\label{stok} Energy dependent $\ord{(g^2)}$ 
contribution to the electron self energy.}
\end{figure}

The graph of Fig. \ref{stok} gives
\bea
I_{\alpha2}&=&\int dxdtdydt'iG_0(x_1,t_1;x,t)iG_0(x,t;y,t')iG_0(y,t';x_2,t_2)
iG_\alpha(x,t;y,t')iG_\alpha(y,t';x,t)\nonu
&=&{1\over a^2_s}\int dx\int dt\int dt'iG_0(x_1,t_1;x,t)
iG_0(x,t;x,t')iG_0(x,t';x_2,t_2)\nonu
&&\times\left[\theta(t-t')\left(1-n_\alpha\right)-\theta(t'-t)n_\alpha\right]
\left[\theta(t'-t)\left(1-n_\alpha\right)-\theta(t-t')n_\alpha\right]\nonu
&=&{i\over a^2_s}\int{d\omega_1\over2\pi}\int{d\omega_2\over2\pi}
\int{d\omega_3\over2\pi}\int{d^2p_1\over(2\pi)^2}\int{d^2p_2\over(2\pi)^2}
\frac{e^{ip_1(x_1-x_2)-i\omega_1(t_1-t_2)}}
{(\omega_1-{p_1^2\over2m}+\mu+i\delta_{p_1}^\mu)^2}\nonu
&&\times\frac{1}{\omega_2-{p_2^2\over2m}+\mu+i\delta_{p_2}^\mu} 
\left(\frac{1-n_\alpha}{\omega_3+i\epsilon}+
\frac{n_\alpha}{\omega_3-i\epsilon}\right)\nonu
&&\times \left(\frac{1-n_\alpha}{\omega_3+\omega_2-\omega_1
+i\epsilon}+\frac{n_\alpha}{\omega_3+\omega_2-\omega_1-i\epsilon}\right).
\eea
The integration over $\omega_3$ of the terms between the brackets yields
\be
i(1-n_\alpha)n_\alpha\left(\frac{1}{\omega_1-\omega_2+i\epsilon}-
\frac{1}{\omega_1-\omega_2-i\epsilon}\right)
=2\pi(1-n_\alpha)n_\alpha\delta(\omega_1-\omega_2),
\ee
thus
\be
I_{\alpha 2}={i\over a^2_s}\int{d\omega_1\over 2\pi}
\int{d^2p_1\over(2\pi)^2}\int{d^2p_2\over(2\pi)^2}
\frac{e^{ip_1(x_1-x_2)-i\omega_1(t_1-t_2)}}
{(\omega_1-{p_1^2\over2m}+\mu+i\delta_{p_1}^\mu)^2} 
\frac{1}{\omega_1-{p_2^2\over2m}+\mu+i\delta_{p_2}^\mu}(1-n_\alpha)n_\alpha.
\ee
The integration over $p_2$ gives
\be
\int{d^2p_2\over(2\pi)^2}{1\over\omega_1-{p_2^2\over2m}+\mu+i\delta_{p_2}^\mu}
=\int{d^2p_2\over(2\pi)^2}\left\{P.P.{1\over\omega_1-{p_2^2\over 2m}+\mu} 
-i\frac{\delta_{p_2}^\mu}{(\omega_1-{p_2^2\over2m}+\mu)^2+\delta^2}\right\}.
\ee
where $P.P.$ stands for the principal part and produces only a shift
in the energy. This shift, being divergent as $a_s\to0$ is not
computable without specifying the details of the ultraviolet cutoff.
But the more important imaginary part is finite and can
be decomposed in the following way
\bea
-i\int{d^2p_2\over(2\pi)^2} \frac{\delta_{p_2}^\mu}
{(\omega_1-{p_2^2\over 2m}+\mu)^2+\delta^2}
&=&i{m\over2\pi}\left(\int_0^\mu-\int_\mu^\infty\right)
d\epsilon\pi\delta(\omega_1+\mu-\epsilon)\nonu
&=&i{m\over2}\left\{\left[\theta(\omega_1+\mu)-\theta(\omega_1)\right]
-\theta(\omega_1)\right\}.
\eea
The term containing $\theta(\omega_1+\mu)-\theta(\omega_1)$ represents the
contribution of the Fermi sphere. The last term stands for the
contributions of the excitations above the Fermi surface.
Their sum is proportional to the free electron density of states,
\be
\tilde{N}_0(E)={m\over2\pi\hbar^2}\left(2\theta(E-\mu)-\theta(E)\right)
\ee
which is positive when $E$ is above the Fermi level and negative for
$0<E<\mu$. The final result
\be
Im I_2(\omega,p)={i\over a^2}\hat{\rho}(1-\hat{\rho})
\frac{1}{(\omega-{p^2\over2m}+\mu+i\delta_{p}^\mu)^2}i{m\over 2} 
\left(\theta(\omega+\mu)-2\theta(\omega)\right).
\ee
gives the imaginary part of the self-energy
\be
Im\Sigma^{(2)}_2(\omega,p)
=-ig^2{m\over2}\rho(1-\hat{\rho})\left(2\theta(\omega)-
\theta(\omega+\mu)\right).
\ee

\end{appendix}
\end{document}